\documentstyle [psfig,11pt]{article}
\begin{document}

\begin{center}
{\Large Imaging in turbid media using quasi-ballistic photons} \\
\vspace*{0.4cm}
Venkatesh Gopal$^{1}$,
Sushil Mujumdar$^{2}$, Hema Ramachandran${^2}$ and A.K.Sood.$^{1,3}$ \\
\vspace*{0.2cm}
${^1}$Department of Physics, Indian Institute of Science, Bangalore
560 012, INDIA\\
${^2}$Raman Research Institute, Sadashivanagar, Bangalore 560 080,
INDIA \\
${^3}$Jawaharlal Nehru Centre for Advanced Scientific Research, \\
Jakkur Campus, Bangalore 560 066, INDIA
\end{center}

\begin{abstract}
We study by means of experiments and Monte Carlo simulations, the
scattering of light in random media, to determine the distance upto
which photons travel along almost undeviated paths within a scattering
medium, and are therefore capable of casting a shadow of an opaque
inclusion embedded within the medium. Such photons are isolated by
polarisation discrimination wherein the plane of linear polarisation
of the input light is continuously rotated and the polarisation
preserving component of the emerging light is extracted by means of a
Fourier transform. This technique is a software implementation of
lock-in detection. We find that images may be recovered to a depth
far in excess of what is predicted by the diffusion theory of photon
propagation. To understand our experimental results, we perform Monte
Carlo simulations to model the random walk behaviour of the multiply
scattered photons. We present a new definition of a diffusing photon
in terms of the memory of its initial direction of propagation, which
we then quantify in terms of an angular correlation function. This
redefinition yields the penetration depth of the polarisation
preserving photons. Based on these results, we have formulated a
model to understand shadow formation in a turbid medium, the
predictions of which are in good agreement with our experimental
results.
\end{abstract}

\section{Introduction}

Photons travelling through turbid media such as milk or clouds are
multiply scattered. The turbidity of such media increases with either
a high refractive index contrast between the scatterers and the
medium, or a large number density of scatterers, or both. What does
it mean to be able to `see' through such a medium?. By the process of
`seeing', one refers to the collection of `image bearing' photons by
the eye, photons that have emerged, despite being scattered, after
travelling nearly undeviated from the direction in which they entered
the medium. The number of such photons depends on the number density
and scattering anisotropy of the scatterers. With very high scatterer
concentrations, one has a situation where the number of image bearing
photons is nearly negligible and most of the photons have little or no
memory of their initial direction of propagation. The transport of
the photon density in such a highly multiply scattering regime, can be
accurately modelled by a diffusion equation. Measurements of this
diffuse intensity can be used to extract useful dynamical information
about the medium as in diffusing wave spectroscopy (DWS) \cite{Pine},
or to detect static and dynamic inclusions within the medium
\cite{Boas_ct, Heckmeier, den Outer}.

Most methods of selecting such image bearing photons rely on
time-gated detection, which utilises the fact that image bearing
photons are the first to emerge from the medium while diffusing
photons travel much longer paths and emerge later \cite{Time_gate}.
However, photons also carry information in the form of their state of
polarisation and the image bearing photons may be discriminated from
those that have travelled longer paths by detecting their
polarisation. The initial state of polarisation is preserved for
photons travelling nearly straight line trajectories while it is
randomised for diffusing photons that have traversed the medium by a
random walk. This fact has been recognised and experiments using a
variety of methods to tag these photons \cite{Alfano1, Alfano2,
Morgan, Schmitt} have been performed. In a recent experiment
\cite{Emile}, light passing through a rotating linear polariser was
made incident on a turbid medium. The emerging scattered light, after
passing through a fixed analyser, was collected by a detector by
scanning across the exit face of the sample. Two pinholes, one before
and another after the sample, permitted predominantly the image
bearing photons to be detected. The resultant signal consisted of an
oscillatory component due to the photons that still retained a
significant amount of their original state of polarisation, riding on
a constant background arising from the multiply scattered and
completely depolarised light. Using lock-in detection to collect only
the oscillatory component, it was possible to image millimeter sized
objects immersed in milk. In a subsequent experiment
\cite{Ramachandran}, the input and exit pinholes were replaced by a
lens-aperture system, and the point detector by a CCD camera. The
lens aperture system rejects much of the diffuse light while the CCD
array permits simultaneous two-dimensional data collection. The
frequency lock-in was achieved by means of software and images of
objects hidden in turbid media were obtained with high resolution
without having to perform multiple scans across the object as in
\cite{Emile}.

Before we proceed further, we provide a brief introduction to multiple
scattering theory to describe the relevant quantities and notation.
Photons entering a scattering medium travel exponentially distributed
ballistic pathlengths between scattering events. The scattering mean
free path $l_{s}$ is the mean distance between scattering events and
is determined by the scattering cross section $\sigma$ and the number
density $\phi$ of the scatterers as $l_{s} = 1/\sigma \phi$. The
scattering cross sections and the probability to scatter at a given
angle are calculated by the well known Mie theory \cite{Bohren and
Huffman}. However, scattering does not always randomise the direction
of the photon. Typically, particles whose sizes are large or
comparable to the wavelength of the incident light cannot be thought
of as simple dipole scatterers and Mie theory shows that the
scattering is strongly peaked in the forward direction (See
\cite{Feynman} for a simple description of why this must be so). The
result of this anisotropy in scattering is that the photon is often
not randomised after a single scattering event and there is a
`persistence length' over which the photon travels, on average, in
approximately the same direction before being randomised. If the
photon undergoes a very large number of scattering events, then one
may assume that the photon performs a random walk and that the photon
flux is transported diffusively within the medium. The persistence
length or the length scale over which the photon is randomised is
called the transport mean free path $l^{*}$ \cite{lstar}. The
`optical thickness' or optical density $\tau$ of a slab of thickness
$L$ is defined as $\tau = L/l^{*}$. When $L \sim l^{*}$ the
scattering is largely ballistic, and, when $L \gg l^{*}$, the
diffusion approximation is valid \cite{Pine,lstar}. The scattering
anisotropy $g$, a measure of the persistence length, is defined in
terms of the mean free paths $l^{*}$ and $l_{s}$ as $g = 1 -
(l_{s}/l^{*})$.

Photons travelling through a random medium may be classified into
three types:  a ballistic component that has not undergone any
scattering, a diffuse component that is completely randomised
directionally and may be modelled by a diffusion equation, and a
quasi-ballistic or `snake' component, that has undergone more than one
scattering event but is still travelling in approximately the same
direction as it did when it entered the medium. The ballistic and
snake photons together form the image bearing component of the
transmitted intensity. The ballistic transmitted intensity falls off
exponentially with the thickness $L$ of the scattering medium as
$I_{ballistic} = \exp(-L/l_{s})$ and is usually too small to detect at
large optical densities. However images with high resolution may
still be acquired using the snake photons. The snake component is
intermediate between the ballistic and diffuse regimes and is
currently poorly understood. This crossover from one regime to
another has often been modelled in an approximate manner by assuming
that a delta function source of diffusing photons lies at a fixed
depth within the medium. This approximation yields excellent results
when the detector collects only diffuse light but fails rather
significantly otherwise \cite{Durian1,Durian2}. Our motivation in the
present work is twofold. First, we seek to understand how far one can
`see' into a random medium using snake photons, and to place limits on
the depth to which such snake photon imaging can be performed. The
second is to investigate the crossover from the ballistic to the
diffusive regime, and thus to understand the characteristics of the
quasi-ballistic photons.

\section{Experimental details}

\subsection{Apparatus}

A schematic diagram of the experimental apparatus is shown in Fig.1.
Light from a 2.5 mW, randomly polarised Helium-Neon laser
(Melles-Griot, $\lambda$=612 nm) is expanded and collimated. The
laser beam passes through a polariser {\bf RP} that is coupled to a
stepper motor driver which rotates the polaroid through 360$^{o}$ in a
variable number of steps. The central part of the expanded and
collimated beam is selected by an iris diaphragm and made incident on
a metal mask {\bf MM} with a 3 mm diameter hole cut into it. Thus, a
circular spot of light of nearly uniform intensity is incident on the
cuvette {\bf S} containing a colloidal suspension. The cuvette has an
internal thickness of 10mm. After passing through the scattering
medium, the transmitted light is collected by the lens {\bf L1}. The
light passing through the lens is apertured by another iris diaphragm
{\bf PH}, which rejects a large fraction of the diffuse flux that
would otherwise be incident on the CCD camera. The light transmitted
through this aperture then passes through a fixed analysing polaroid
{\bf FA} and is incident on an intensified, variable gain CCD camera.
The camera has an 8 bit (256 grayscale values) intensity resolution
with a 512 $\times$ 512 pixel array.

\begin{figure}[!htbp]
\centerline{\psfig{figure=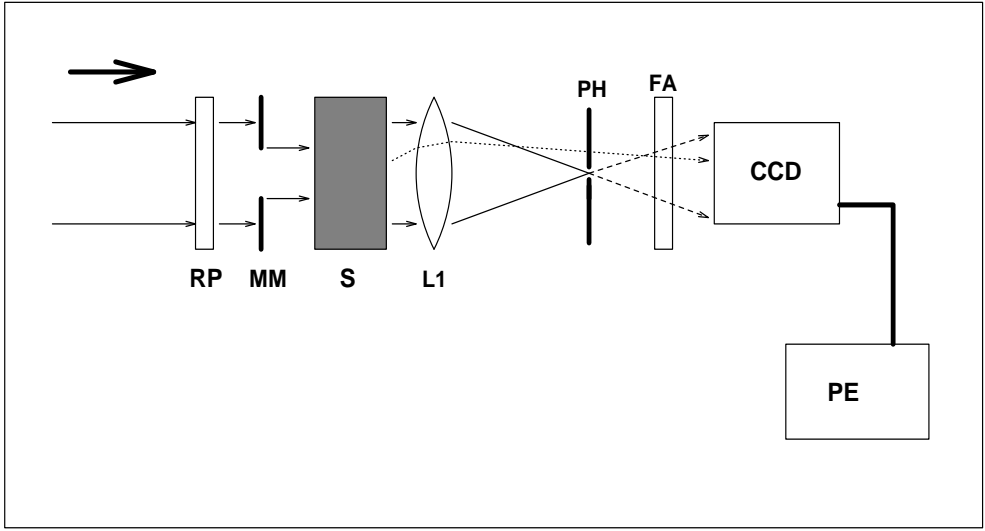,width=13cm,height=7cm}}
\caption{Schematic diagram of the experimental apparatus. {\bf RP} :  Rotating
polaroid, {\bf MM} :  Metal mask with 3 mm diameter hole, {\bf S} :
Cuvette containg scattering medium, {\bf L1} :  Lens ($f$ = 90 mm), {\bf
PH} :  Aperture , {\bf FA} :  Fixed analyser, {\bf CCD} :  Intensified
CCD camera with variable gain, {\bf PE} :  Processing electronics.}
\label{setup}
\end{figure}

It can be seen from Fig. 1 that in the absence of the colloidal
suspension, parallel rays of the incoming beam, on passing through the
lens {\bf L1}, converge to a focus at the aperture {\bf PH} and then
diverge to form a spot on the CCD. The size of the spot is determined
by the focal length of {\bf L1} and the distance of the CCD from {\bf
PH}. On interposing the colloidal suspension, lens {\bf L1} then
images not only the snake photons that emerge normal to the face of
the cuvette, but also some fraction of the diffusively scattered light
that emerges along paths parallel to the snake photons. Since, in the
absence of the pinhole diffuse photons that emerge at angles not
normal to the cuvette would also be imaged onto the CCD array, it was
necessary to limit the acceptance angle of the detector with an
aperture. The dotted line in Fig. 1 shows one such a photon
trajectory which would be incident on the detector if it were not
rejected by the aperture {\bf PH}.

The experiment is performed as follows. First, a direct image without
the sample is recorded to serve as a reference. A sample image
obtained is shown in Fig.2. The concentric diffraction rings seen in
the image are due to diffraction from the edges of the mask {\bf MM}.
The dark region at the top right of the bright spot of light, is due
to a defect on one of the optical components. The cuvette with the
colloidal suspension is now introduced into the beam. To ensure that
the maximum dynamic range of the detector is used, the polarisers are
crossed and the gain is increased such that, on making the two
polaroids parallel, the pixels in the brightest region of the image
are almost saturated. However, as the diffuse flux begins to increase
at high optical densities, the baseline intensity, the intensity when
the polarisers are crossed, is not zero any longer. The amplitude of
the intensity oscillations rapidly decreases beyond this point as the
snake photon population is sharply reduced and the diffuse flux begins
to dominate the scattered intensity. The polariser RP is rotated in
steps of 15$^{o}$ and at each step an image is recorded and saved.
The process is continued until 512 frames are recorded. Experiments
were performed for each of the colloidal suspensions of the following
optical densities :  $\tau$ = 0.5, 1.1, 2.5, 3.0, 3.6, 5.2, 5.9, 6.7,
8.8, 9.5, 10.5, 11.6 , and 12.6, which were prepared using 0.23$\mu m$
diameter polystyrene particles (Bangs Laboratories, U.S.A.)  suspended
in water. At $612 nm$, these particles were calculated to have a
scattering anisotropy, $g = 0.45$.
\begin{figure}[!htbp]
\centerline{\psfig{figure=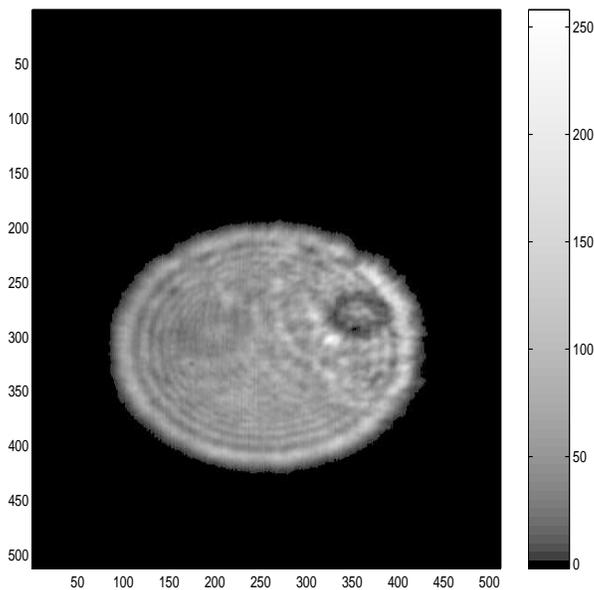,width=8cm,height=8cm}}
\caption{Direct image of the laser beam recorded without the scattering medium.}
\label{dir-beam}
\end{figure}
\subsection{Data Processing}

Figure 3 shows in schematic form, the procedure for processing the
images. At the end of each run, we obtain a deck of 512 frames
ordered in time. Consider any pixel from an image of this deck, say
the pixel $P_{k}(i,j)$. The indices $i$ and $j$ refer to the location
of the pixel within the $k$th image of the deck. A time series is
formed for the $(i,j)$th pixel by storing all the time ordered values
that this pixel takes as the polaroid {\bf RP} is rotated. This time
series is then Fourier transformed using a Fast Fourier Transform
(FFT) routine which yields two peaks, one at zero frequency, and
another at the rotational frequency of the polaroid, $\omega =
\pi/N_{s}$ ($N_{s}$ = 12, the number of images in one half rotation of
the polaroid, after which the signal repeats itself). The amplitudes
of the zero frequency peak and that at the rotational frequency are
stored and two separate images are built up. In one image, termed as
the `zero frequency' image, the pixel with coordinates $(i,j)$ is
assigned the value of the amplitude of the peak at zero frequency. In
the other image, the pixel with coordinates $(i,j)$ is assigned the
amplitude of the peak obtained at the rotation frequency. This image
is termed the `polarisation retaining' image.

\begin{figure}[!htbp]
\centerline{\psfig{figure=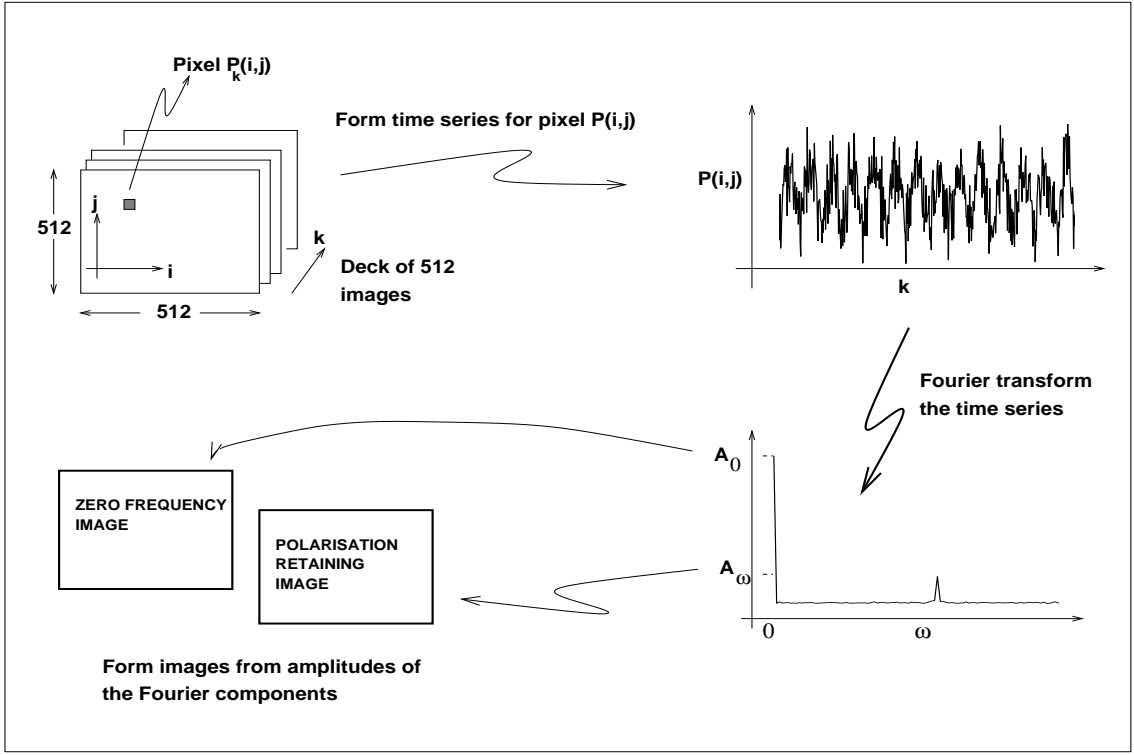,width=15cm,height=10cm}}
\caption{Schematic representation of the data processing procedure. ${\bf
A_{0}}$ and ${\bf A_{\omega}}$ are the Fourier amplitudes at the zero
and the polaroid rotational frequencies respectively.}
\label{dataproc}
\end{figure}

\section{Experimental results}

We have imaged the bright spot of light that is transmitted by the
metal mask {\bf MM} through colloidal suspensions of $0.23 \mu m$
particles of a wide range of optical densities. While a direct image
was seen in the optically thin samples ($\tau < \sim 7$), no image was
directly discernible at larger $\tau$. However, on processing the
data as described earlier, the circular spot of light was visible in
the polarisation retaining image and images could be extracted upto
$\tau \sim 10$. Figure 4 shows a series of images obtained at optical
densities of 5.94 (figs. a \& b), 8.76 (figs. c \& d) and 9.54
(figs. e \& f). The images on the left are the polarisation
retaining images , while those on the right are the zero frequency
images. Figures 4(c-f) have been smoothened by applying a 10 $\times$
10 median filter to improve the contrast and remove some of the noise
.In figs. 4(e) and 4(f), the camera has been moved closer to the
aperture and so the bright transmitted spot of light is smaller than
in the previous images. These images are recorded at high gain levels
of the CCD camera and are noisy because the CCD array is extremely
sensitive to stray light at high gain. That we have not been entirely
successful in cutting off all stray light is evident from the bright
streaks that can be seen at the top and bottom of some of the images
.Figures 4(c) and 4(e) clearly bring out the efficacy of the technique
.While the zero frequency image 4(d) faintly shows a part of the
outline of the bright disc, and 4(f) shows no details at all, a clear
image is visible in the corresponding polarsation retaining images
4(c) and 4(e).
\begin{figure}[!htbp]
\centerline{\psfig{figure=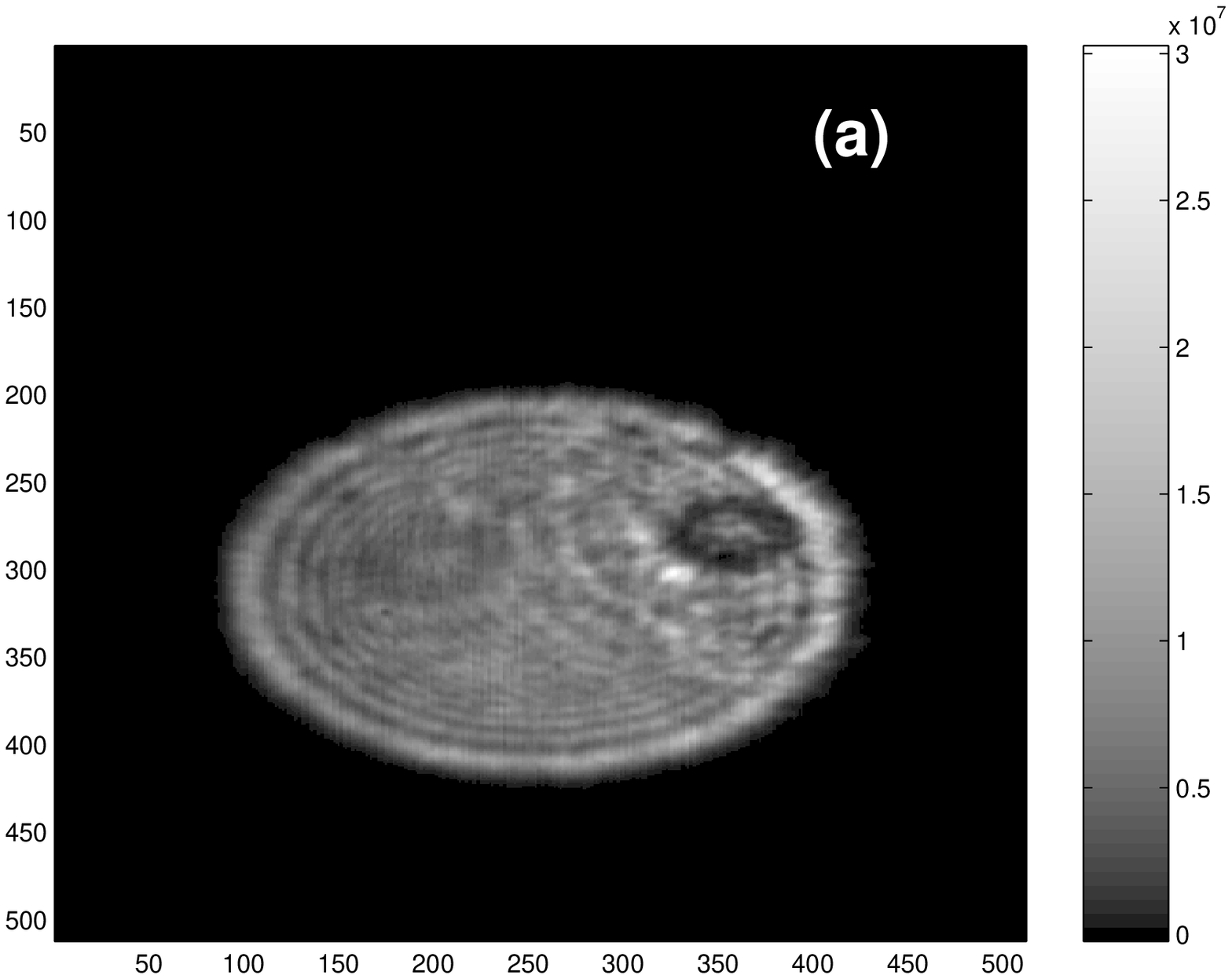,width=7.2cm,height=7.2cm}
\psfig{figure=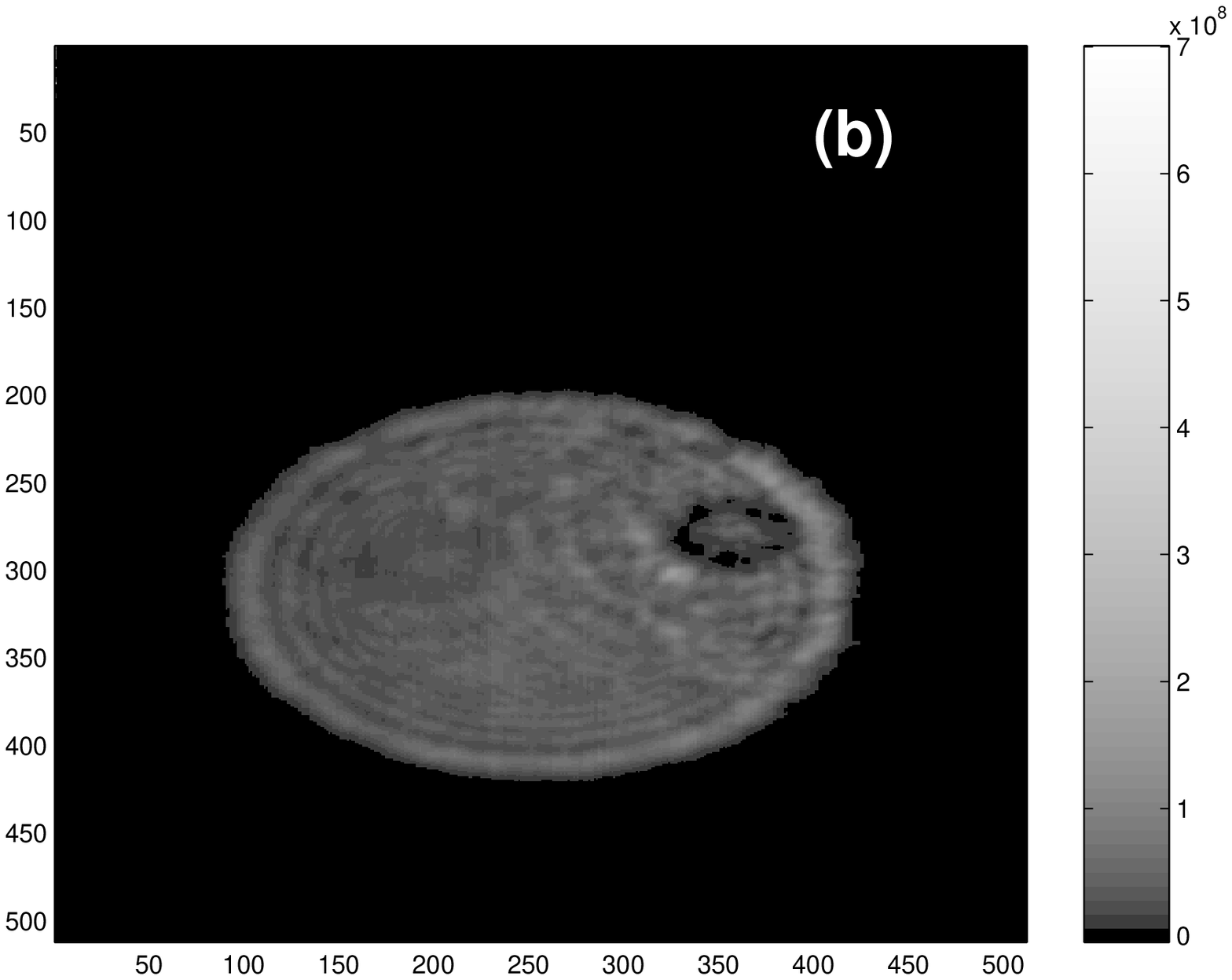,width=7.2cm,height=7.2cm}}
\centerline{\psfig{figure=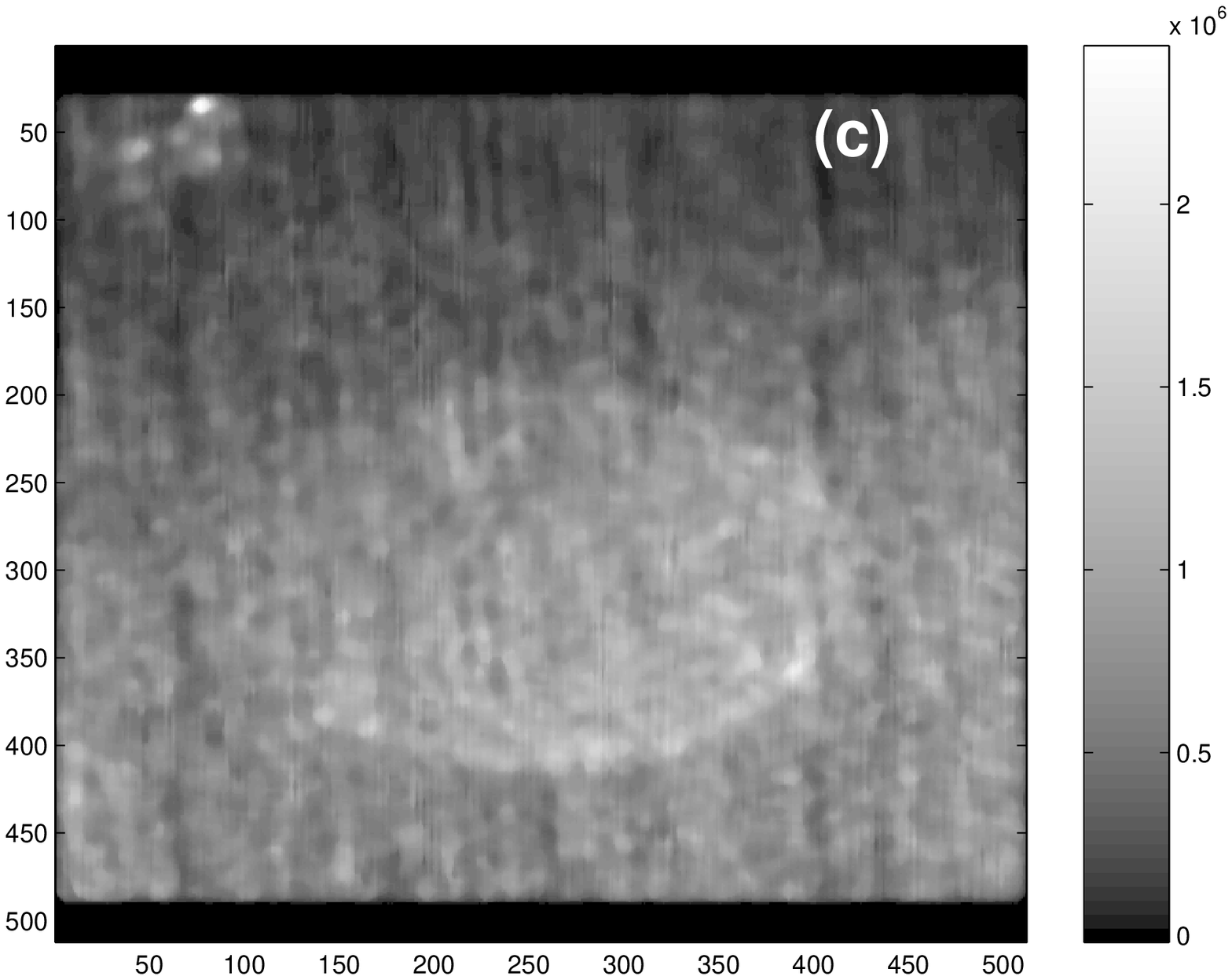,width=7.2cm,height=7.2cm}
\psfig{figure=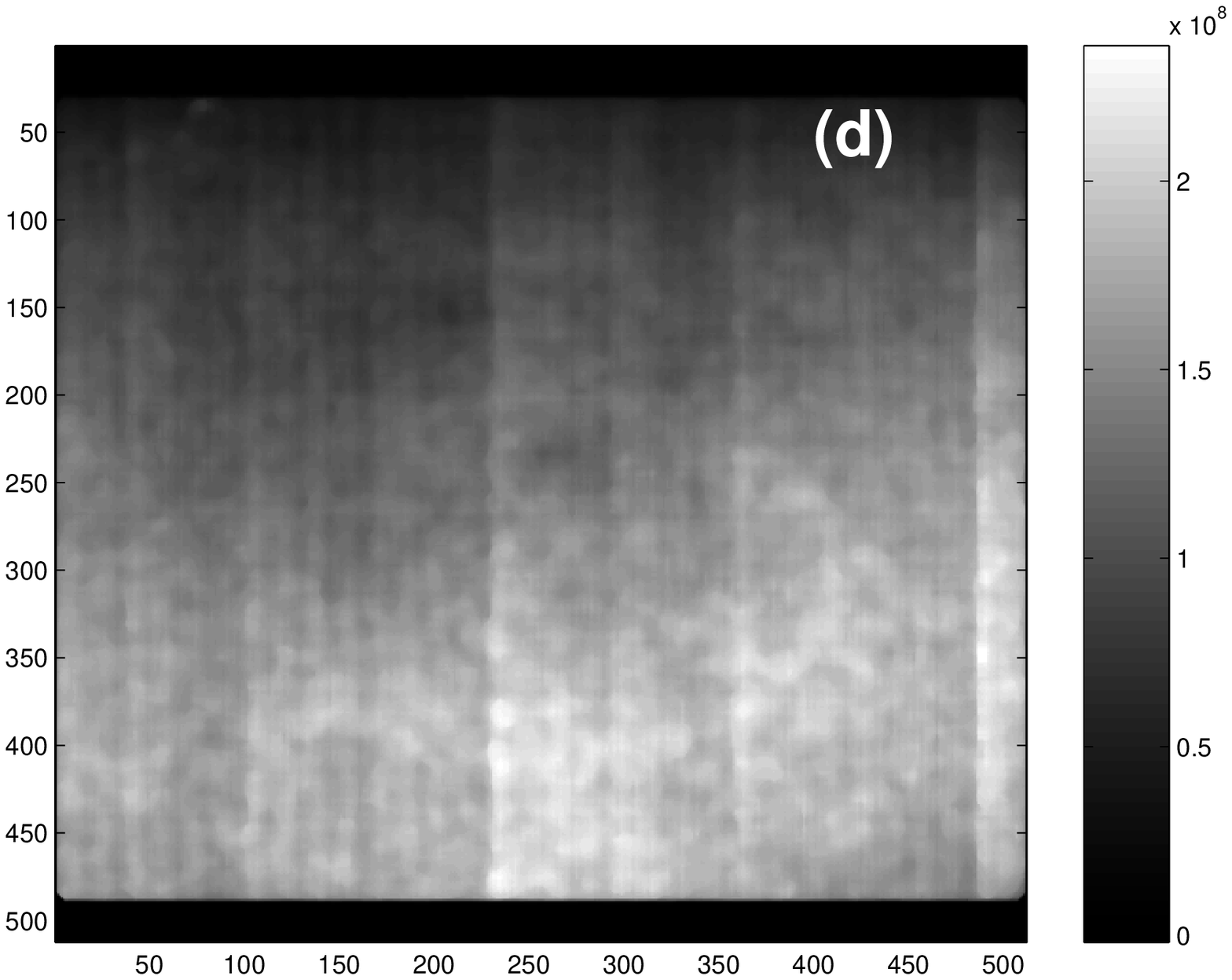,width=7.2cm,height=7.2cm}}
\centerline{\psfig{figure=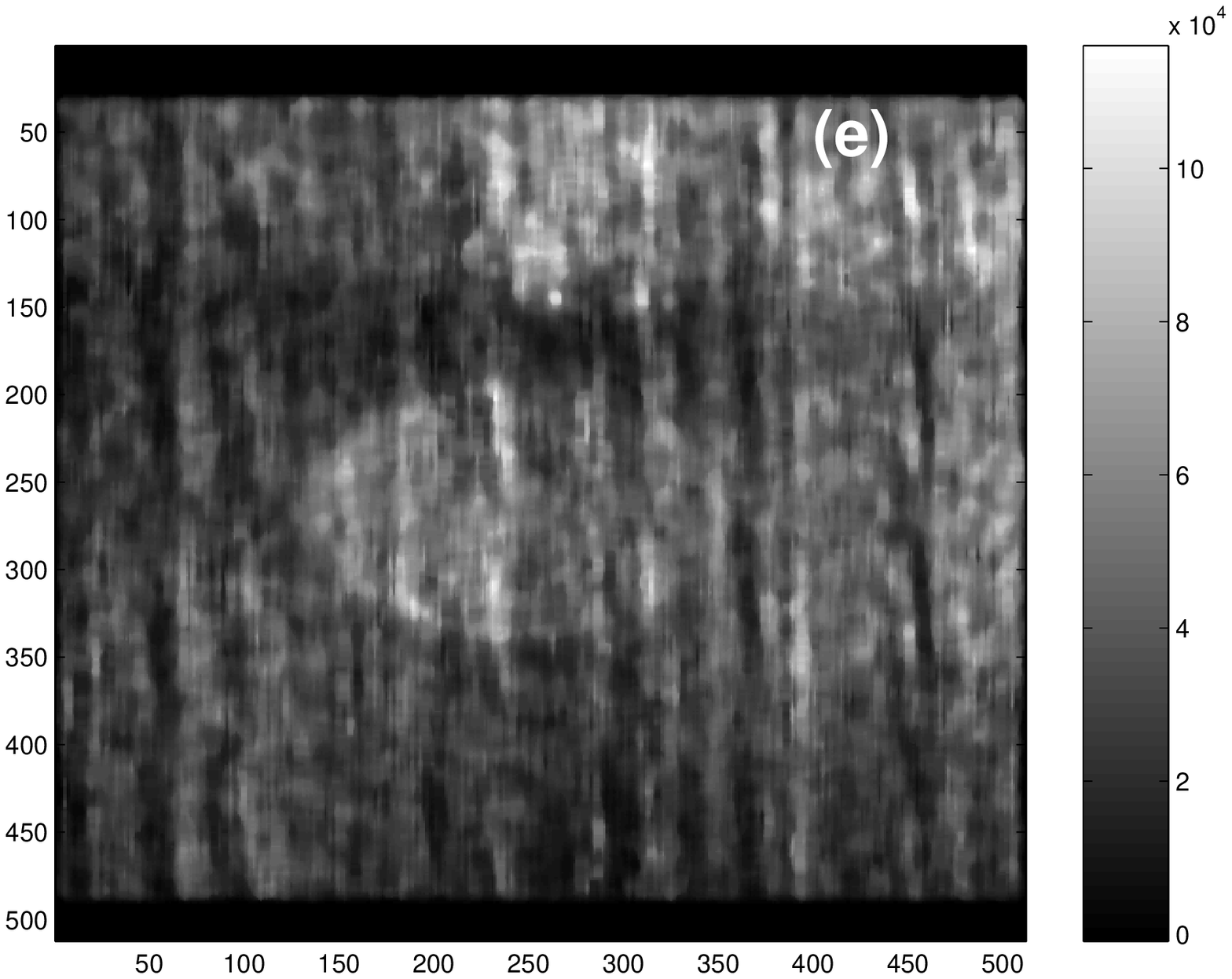,width=7.2cm,height=7.2cm}
\psfig{figure=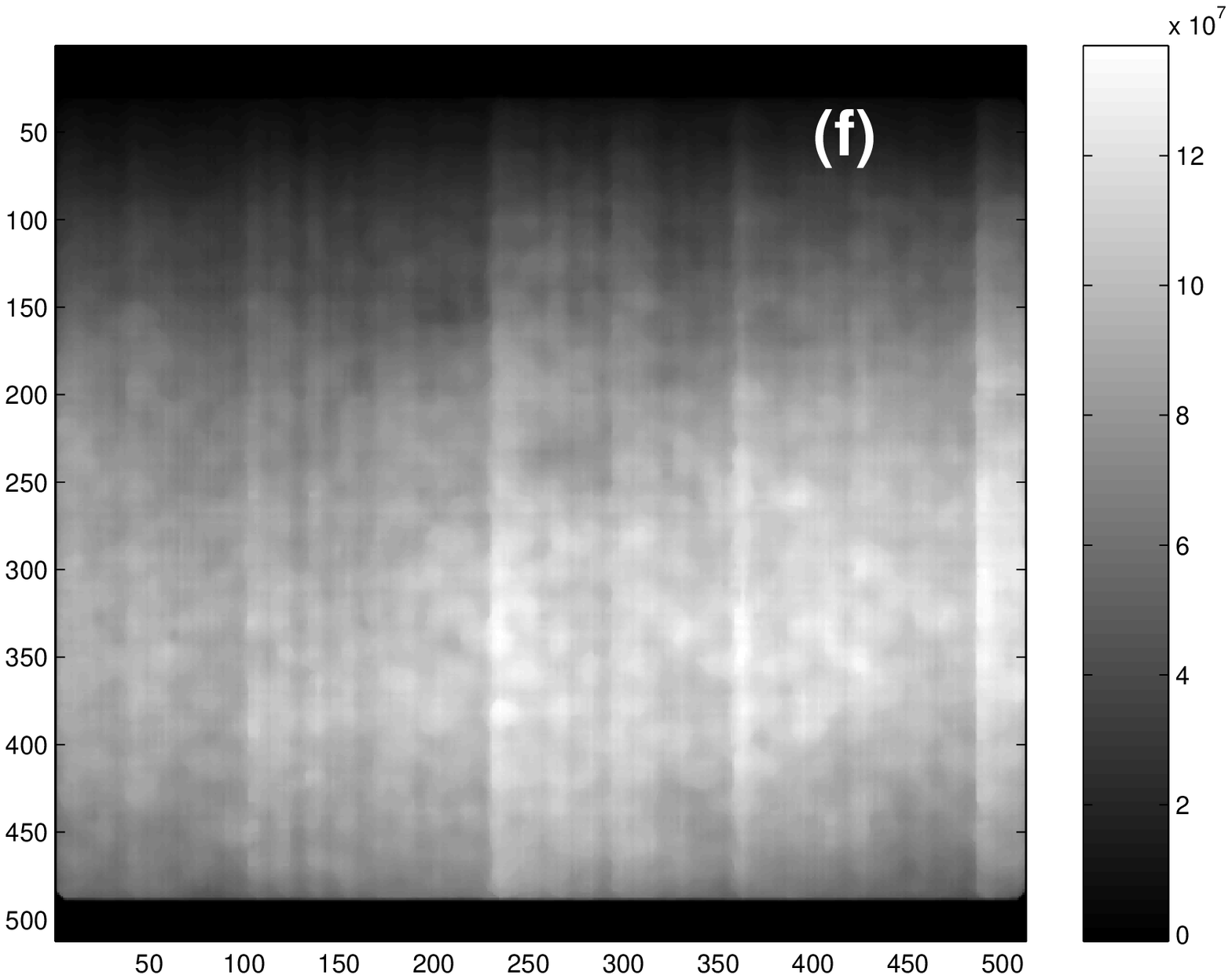,width=7.2cm,height=7.2cm}}
\caption{Images recorded at the following optical densities (a,b) $\tau = 5.94$,
(c,d) $\tau = 8.76$ and (e,f) $\tau = 9.54$. The polarisation
preserving images ((a),(c), and(e)) are in the column on the left while
the zero frequency images ((b),(d), and(f)) are in the one on the right.}
\label{images}
\end{figure}

We have studied the variation of the quality of the images obtained as
a function of the optical density of the scattering medium. A number
of parameters could be used to quantify the quality of an image, such
as the sharpness of the edges, or the visibility of features within
the image. However, we are interested in determining the fraction of
the transmitted photons that preserve their polarisation, as a
function of the optical density of the medium. Therefore, we measure
for each pixel, the ratio of the Fourier amplitudes ${\bf R} =
A_{\omega}/A_{0}$. Here $A_{0}$ is the amplitude of the zero
frequency component, and $A_{\omega}$ the amplitude of the rotational
frequency component obtained on Fourier transforming the time series
for a given pixel. The ratio ${\bf R}$ is calculated and averaged for
each pixel within a region of approximately 200 $\times$ 200 pixels,
which is chosen at the centre of the circular transmitted spot of
light. The pixels in our camera are rectangular causing the circular
spot of light transmitted by the mask to appear oval in the images.
Since the shape of the object is not important to our analysis, we do
not correct for this distortion of the aspect ratio.

To understand the parameter {\bf R} better, consider two limiting
cases of imaging, one in which the medium is transparent with no
scattering, and at the other extreme, one in which the entire
scattered flux is diffuse and the number of quasi-ballistic photons is
negligible. The scattered light may be thought of as consisting of a
polarised component whose intensity varies as $\cos^{2}\theta$, where
$\theta$ is the relative angle between the rotating polaroid and the
fixed analyser, and a randomly polarised diffuse component of constant
intensity. The signal in the case of a transparent medium consists of
only an oscillatory component. The ratio of the Fourier amplitudes in
this case would then be given by ${\bf R}=0.5$ as can be seen from our
choice of definition of the Fourier coefficients \cite{Fourier_defn}.
In other words, for a purely oscillatory signal, the mean is half the
peak-to-peak swing of the signal. In the case of a highly turbid
medium, we have a constant intensity at any orientation of the
rotating polaroid and ${\bf R}=0$.

\begin{figure}[!htbp]
\centerline{\psfig{figure=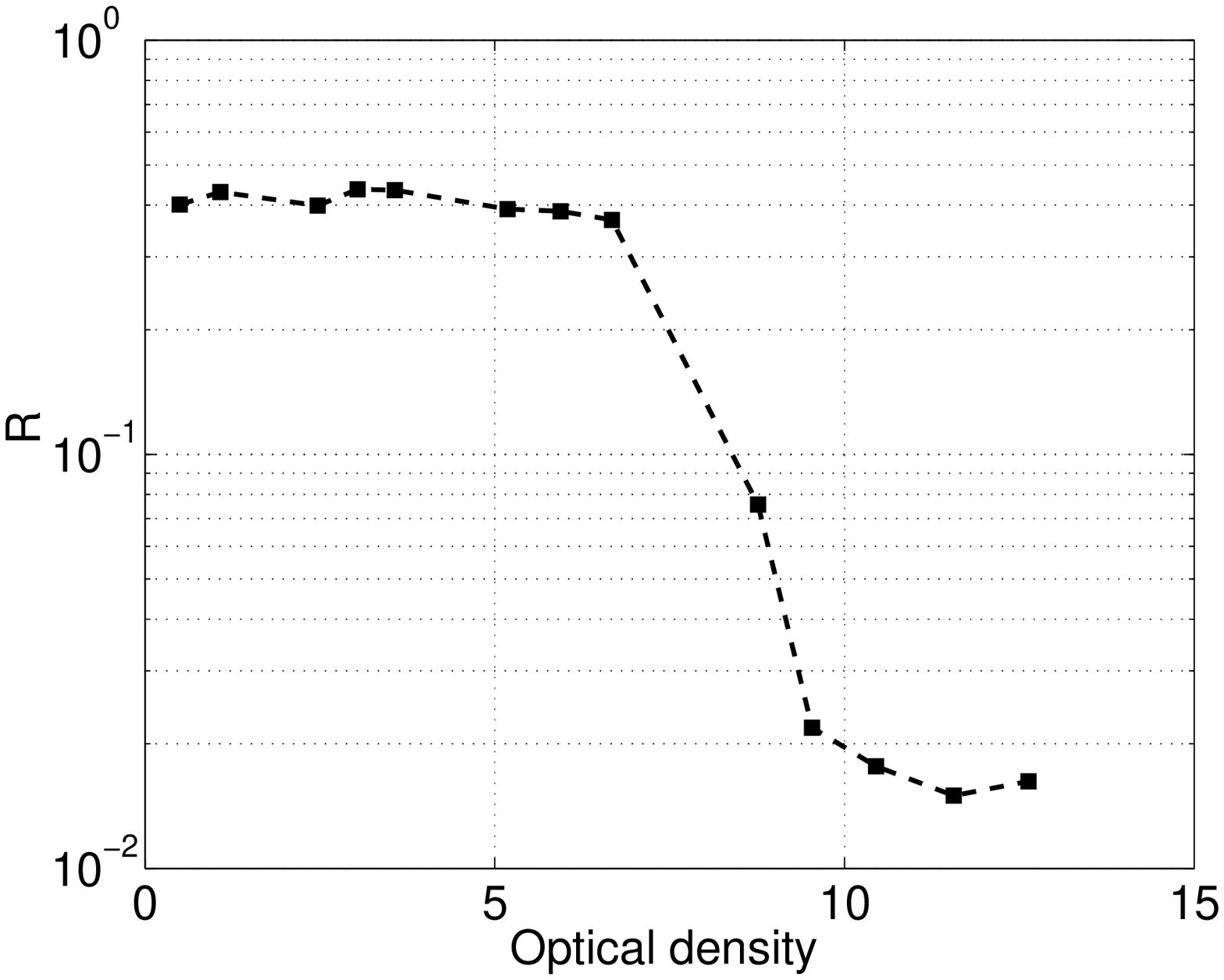,width=9cm,height=9cm}}
\caption{Variation of the image visibility parameter ${\bf R} =
A_{\omega}/A_{0}$, as a function of the optical density $\tau =
L/l^{*}$.}
\label{rparam}
\end{figure}
Figure 5 shows the variation of ${\bf R}$ with increasing optical
density $\tau$. We observe that upto $\tau \sim 7$ there is almost no
variation in ${\bf R}$ and it is also very close to the maximum value
of 0.5 that can be attained for a signal consisting of {\it only} the
time varying component. For the range $7 < \tau < 10$ we observe an
exponential decay in R, while beyond $\tau >10$ the ratio remains
constant and no images could be extracted. These results are rather
surprising for two reasons. First, according to the diffusion model
commonly employed, it is assumed that photons are randomised on a
length scale comparable to $l^*$. In fact, in deriving an expression
for the diffuse flux that crosses a given area element in a random
medium, it is assumed \cite{Ping Sheng}, that the probability for a
photon to travel a path length $s$ without being randomised is given
by $P(s) = \exp(-s/l^{*})$. Thus diffusion theory would predict that
the non-randomised or image bearing flux at a depth of 7$l^{*}$ would
be only a fraction $\sim 9 \times 10^{-4}$ of the incident flux. As
we show later, this fraction would be comparable to the diffuse
background intensity, making $R$ much less than 0.5. Secondly, one
would expect a gradual decrease in ${\bf R}$ and not the abrupt change
that is observed since there is a smooth conversion of photons from
quasi-ballistic to diffusive transport.

\section{Discussion}

Our aim now is to explain our results within the framework of a simple
model that would provide order-of-magnitude estimates of the
visibility of images obtained through random media. On the one hand,
when the scattering is minimal ($\tau < \sim 3$), the analysis is
simple, and on the other, in the highly multiple scattering limit too,
solutions can be found by a suitable application of the diffusion
equation. However, the nearly constant value of ${\bf R}$ upto $\tau
\sim 7$ urges us to reconsider our understanding of what constitutes a
`diffuse' photon. In this section we describe results obtained by
Monte Carlo random walk simulations of photon transport in scattering
media. These simulations are central to the understanding of our
experimental results.

\subsection{Monte Carlo simulations}

The procedure for our Monte Carlo simulations was as follows. Photons
were launched from the centre of an infinite slab. The simulation
modelled photon transport assuming that the photons travel
exponentially distributed lengths $s$ between scattering events. The
probability $P(s)$ of travelling a ballistic path length $s$ is given
by the familiar Lambert-Beer law, $P(s) = exp(-s/l_s)$, where $l_s$ is
the scattering mean free path of the photons in the medium. The
random paths between scattering events were generated taking $s = -l_s
\cdot \ln(RAN)$, where RAN is a random number uniformly distributed
between 0 and 1 \cite{Numrec}. The scattering angles were chosen such
that they had a distribution of directions given by the
Henyey-Greenstein phase function \cite{Durian1}, where the probability
of scattering at an angle $\theta$ relative to the incident direction
of the photon is given by \begin{equation} P(\cos \theta) = \frac{1 -
g^2}{(1 + g^2 - 2g \cos \theta)^{3/2}} \end{equation}

\subsection{Defining a diffusing photon}

In the spirit of diffusion theory, we assume that a diffuse photon is
one which has lost all memory of its initial direction of propagation.
The scattering anisotropy is a measure of the ability of a photon to
`remember' its initial direction after a single scattering event. For
a scattering anisotropy $g = 1$, the photon always travels along its
initial direction of propagation, while for $g = 0$, it is randomised
at each scattering event. We have used random walk simulations to
investigate the variation of the number of scattering events that it
takes for the direction of the incident photon to be completely
randomised, as a function of the scattering anisotropy. Photon
trajectories were generated, as described previously, in an infinite
medium. The directional correlation function $C(j) = \langle {\bf
\hat{n}}(0) \cdot {\bf \hat{n}}(j) \rangle$, where ${\bf \hat{n}}(j)$
represents the unit vector along the photon trajectory after the {\it
j}th scattering event, was calculated for eight different values of
$g$, averaging over $10^{4}$ trajectories in each case. A plot of
$C(j)$ is shown in Fig. 6 for three values of $g$. At long times,
$C(j)$ fluctuates about zero. We calculate the standard deviation in
the fluctuations of $C(j)$ at long times and then find the point where
$C(j)$ drops below this value for the first time. This value of $j$
is taken as the number of scattering events ${N_{d}}$ required for a
complete loss of angular correlation. It can be seen from Fig. 6,
that for isotropic scatterers it takes only a few ($\sim 3$ for $g =
0.06$) events for angular memory to be lost, while on the other hand,
the anisotropic scatterers require many more scattering events ($\sim
$ 9 for g = 0.42 and $\sim $ 17 for $g = 0.72$). This however, yields
no information about the length scale over which the photons undergo
${N_{d}}$ scattering events, and naturally, begs the question, how far
away from the point at which it is launched, is a photon, when it
loses angular memory?  We answer this question in the next section.
\begin{figure}[!htbp]
\centerline{\psfig{figure=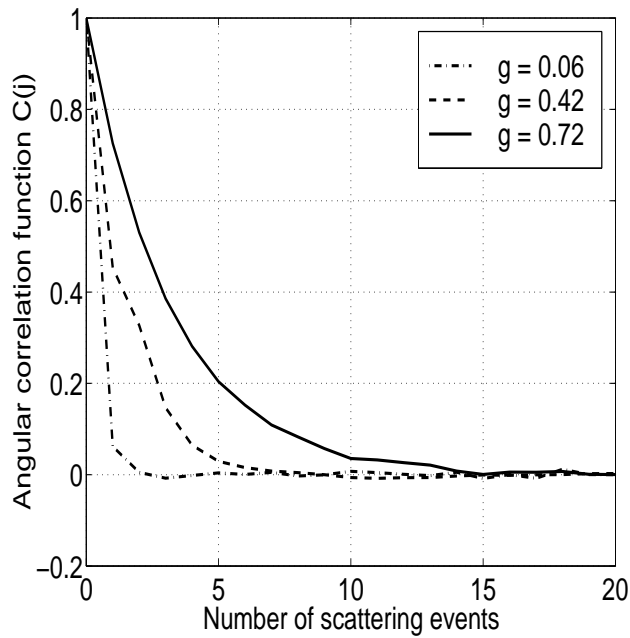,width=9cm,height=9cm}}
\caption{The angular correlation function $C(j) = \langle {\bf \hat{n}}(0) \cdot
{\bf \hat{n}}(j) \rangle$ is shown as a function of the number of
scattering events $j$, for three different values of the scattering
anisotropy $g$.}
\label{corrfun}
\end{figure}

\subsection{Where is a diffusing photon?}

According to diffusion theory, the randomisation of a photon takes
place typically after travelling a distance $l^{*}$, and usually the
source of diffusing photons is modelled by placing a delta function
source at a depth of $l^{*}$ within the medium. However, it is also
well known that this assumption, especially when considering photon
transmission, is inaccurate when the thickness of the scattering
medium is less than about $8l^{*}$ \cite{Kaplan}. Clearly, photons
are not randomised at any one single depth inside the medium and
obviously there exists a smooth distribution of lengths over which the
initially quasi-ballistic flux is converted to a diffusive one.

\begin{figure}[!htbp]
\centerline{\psfig{figure=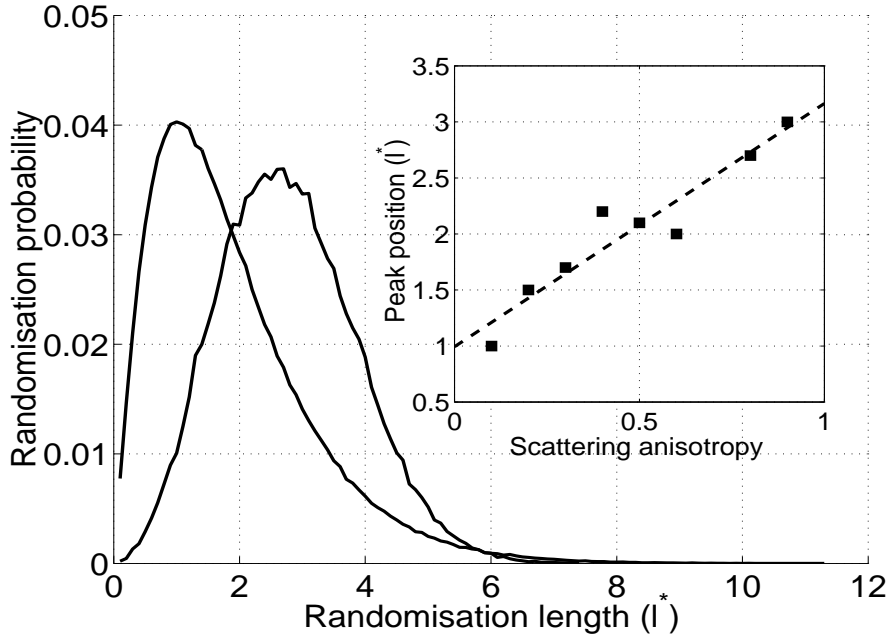,width=12cm,height=9cm}}
\caption{Normalised histograms obtained for the randomisation probability that a
snake photon will undergo $N_{d}$ scattering events, are shown as a
function of the distance travelled from the source. This distance is
termed the `randomisation length'. Curves `a' and `b' are obtained for
scattering anisotropies of 0.1 and 0.8 respectively. The inset shows
the variation of the position of the peak of these histograms as a
function of the scattering anisotropy.}
\label{hist}
\end{figure}

Random walk simulations for the following anisotropy values :  $g = $
0.1, 0.2, 0.3, 0.4, 0.5, 0.6, 0.8 and 0.9, were used to obtain this
distribution of lengths. Once again photons were launched from the
origin, but in a semi-infinite half-space instead. For a given
scattering anisotropy though, we know from the previous simulation,
the number of scattering events $N_{d}$ required for the photon to
lose directional memory. The simulation propagated photons as before
but terminated the trajectory either when the photon had scattered
$N_{d}$ times, or when the photon had been backscattered out of the
half space. Boundary reflections were neglected and absorbing
boundary conditions were applied at the back face of the semi-infinite
medium. The radial distance travelled from the origin after $N_{d}$
scattering events was stored and a histogram of the distribution of
these radial distances was constructed. The histogram bins had a
width of $l^{*}/10$ and were normalised by the total number of photons
that underwent $N_{d}$ scattering events. In the subsequent
discussion, after this normalisation, we treat the number stored in
each bin as the randomisation probability for the snake photons. This
however is strictly not true and we comment on errors introduced by
our choice of normalisation in a later section. It must also be noted
here, that by counting only those photons that undergo $N_{d}$
scattering events within the medium, we have implicitly `defined' a
snake photon as one that has {\it not} left the medium before it has
scattered $N_{d}$ times. Once again, the ambiguity inherent in such a
definition is overlooked in view of the close agreement obtained with
experimental results. Figure 7 shows the distribution of
`randomisation lengths' for scattering anisotropies of 0.1 (curve a)
and 0.8 (curve b) The inset shows the peak position of these
distributions as a function of $g$.

In Fig. 7, two features are to be noted. The distribution is sharper
for the nearly isotropic scatterers, and, becomes more symmetric as
the scattering anisotropy is increased. From the results of the
previous section, we know that for $g = 0.1$, only about three
scattering events are required for a loss of directional memory,
whereas for $g = 0.8$, about 25 scattering events are required. The
symmetry and shape of the peaks may now be understood as a consequence
of the central limit theorem. To do this, reconsider the simulation
as follows. The simulation may be thought of as taking $N_{d}$
vectors from a set of vectors whose lengths are exponentially
distributed as $P(s) = exp(-s/l_{s})$, where $P(s)$ is the probability
of drawing a vector of magnitude $s$. The vectors are then placed
head to tail at angles that are drawn from a set of angles distributed
according to the Henyey-Greenstein function. The histogram is the
distribution of lengths of the resultant displacement vectors. For
nearly isotropic scatterers, $l^{*} \sim l_{s}$, and therefore, for
lengths less than a transport mean free path there is a very small
probability that a photon will be randomised since it is most likely
to be at least one mean free path away from the source after even a
single scattering event. Therefore, when $N_{d}$ is small, the
distribution will be peaked approximately at $s \sim l^{*}$, will
exhibit a sharp fall-off for $s < l^{*}$ and decay more gradually for
$s > l^{*}$. On the other hand, with increasing anisotropy, $N_{d}$
is much larger. As can be seen from Fig. 6, for $g = 0.8$, only the
first 10 or so scattering events are strongly correlated. After that,
only a weak directional correlation persists. Thus, when $N_{d} \sim
10$ or more, the central limit theorem is approximately valid since
the correlations are weak and the resultant distribution tends towards
a Gaussian and a smoother, broader peak is obtained.

The inset in Fig. 7 shows the positions of the peaks of the
randomisation length histograms as a function of the scattering
anisotropy. We see that for nearly isotropic scatterers, the peak
lies near $l^{*}$ and moves out to a maximum of about $3l^{*}$ for a
scattering anisotropy of $0.9$. Thus, we see that the length scale
$l^{*}$ arises {\it naturally} when the scattering is almost isotropic
and corresponds to the position of the peak of the distribution of
randomisation lengths. With increasing anisotropy, the scattering is
peaked in the forward direction and the peak of the distribution lies
further into the medium.

There is no single randomisation length as supposed by diffusion
theory but instead a distribution of lengths over which the conversion
to diffusive transport takes place. According to diffusion theory,
the average number of scattering events that the photon needs to
experience to be randomised is $n = l^{*}/l_{s}$. We compare this in
Table 1, with the number $N_{d}$ obtained from our model and find that
$N_{d} \gg n$. Figure 8 shows the data in Table 1 as a graph.

\begin{center}

\begin{tabular}{|c|c|c|c|c|c|c|c|c|} \hline

Scattering anisotropy &0.1& 0.2 & 0.3 & 0.4 & 0.5 &0.6&0.8 &0.9 \\

($g = 1 - l_{s}/l^{*}$)&&&&&&&& \\ \hline

$n = l^{*}/l_{s}$&1.1&1.25&1.43&1.67&2.0&2.5&5&10 \\ \hline

$N_{d}$&3&5&6&8&12&14&25&53 \\ \hline

\end{tabular}

{\bf Table 1} Comparison of the number of scattering events required
to randomise the photon. $N_{d}$ is calculated from a Monte Carlo
simulation while $n$ is the value predicted by diffusion theory.

\end{center}

\begin{figure}[!htbp]
\centerline{\psfig{figure=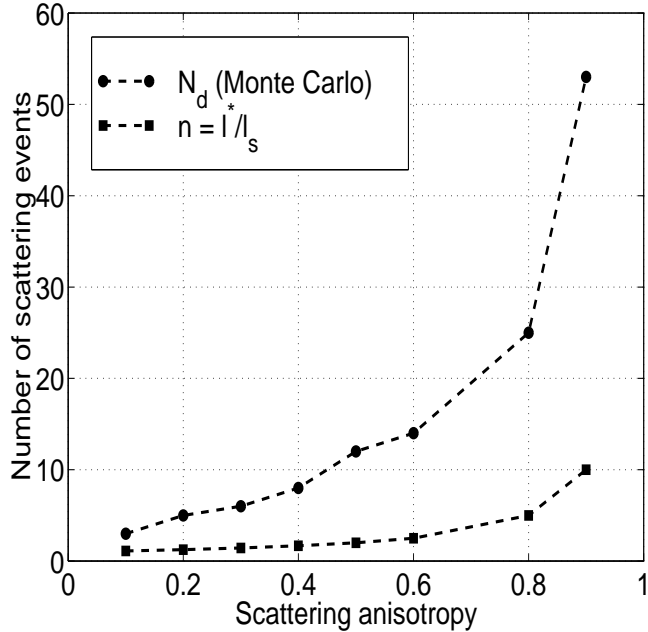,width=9cm,height=9cm}}
\caption{Comparison between $n = l^{*}/l_{s}$ and $N_{d}$, the average number of
scattering events required to randomise the direction of a photon as
predicted by diffusion theory and as obtained by Monte Carlo simulations
respectively.}
\label{steps}
\end{figure}
\noindent

By integrating the area under the curve for the randomisation
probability shown in Fig. 7, we obtain the fraction of the snake
photon flux that is randomised as a function of the distance travelled
from the source. This fraction is denoted by ${\bf I_{diff}}(s)$,
where $s$ is the distance travelled from the source. To simplify the
analysis of our images we make the assumption that a photon retains
its directional memory and polarisation until it undergoes $N_{d}$
scattering events. This component of the flux which retains its
initial polarisation, is the one responsible for the formation of the
polarisation retaining image and is denoted by ${\bf I_{pol}}(s) = 1 -
{\bf I_{diff}}(s)$. Undoubtedly this is an oversimplification but as
we shall see later it yields excellent order-of-magnitude estimates.
Figure 9 shows a semilogarithmic plot of ${\bf I_{pol}}(s)$ as a
function of $s$. The dashed line is the diffusion theory prediction
for the fraction of the incident flux that is not randomised;
$\exp(-s/l^{*})$. The figure shows the effect of the scattering
anisotropy on the rate at which the snake photons are converted to a
diffusive flux. The curves marked `a', `b' and `c' represent
scattering anisotropies of 0.1, 0.4 and 0.8 respectively. At short
distances, due to the strong forward scattering for the highly
anisotropic scatterers, we see that the flux is barely diffused for a
length of $2l^{*}$. For the isotropic scatterers though, the rate of
conversion is much faster. However, irrespective of the value of $g$,
we find that the diffusion approxmation greatly underestimates the
snake photon intensity.

\begin{figure}[!htbp]
\centerline{\psfig{figure=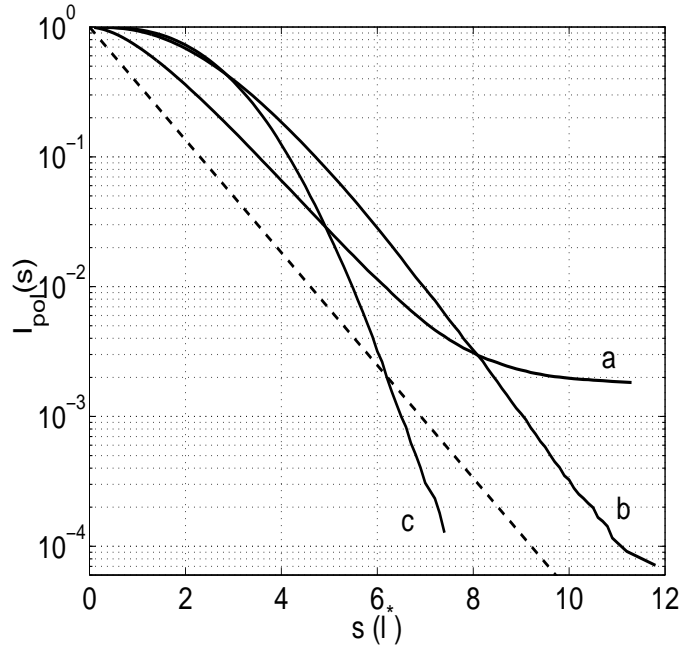,width=9cm,height=9cm}}
\caption{Semilogarithmic plot of ${\bf I_{pol}(s)}$, shown here as a function of $s$,
the depth travelled into the medium, and compared for different
scattering anisotropies $g$. The dashed line is the diffusion theory
prediction for ${\bf I_{pol}}(s)$, while the curves marked `a', `b' and
`c' are obtained for $g =$ 0.1, 0.4 and 0.8 respectively. The depth $s$
is measured in units of $l^{*}$.}
\label{i-pol}
\end{figure}

A more interesting and counterintuitive feature is seen in the tail of
the distributions. After traversing a depth $\sim 3l^{*}$, ${\bf
I_{pol}}(s)$ shows an exponential roll-off for all values of $g$.
However, surprisingly, the tail of curve `a' flattens out around
$8l^{*}$ unlike that of curve `c' for $g = 0.8$. One would have
expected rather that a higher scattering anisotropy would lead to a
longer persistence length. This leads us to the surprising conclusion
that one can image deeper into a turbid medium with isotropic
scatterers than is possible through a medium where the scatterers are
anisotropic. We comment on this point in a more quantitative manner
in the next section.

\section{Putting it all together}

We are now in a position to construct a model that provides
quantitative agreement with our experiments. Consider a collimated
beam of photons entering a slab of scattering medium of thickness $L$.
The fraction of photons that are transmitted, according to the
diffusion approximation, is given by the diffuse transmission
probability $T_{d}$ \cite{Durian3},

\begin{equation} T_{d} = \frac{1 + z_{e}}{(L/l^{*}) + 2z_{e}}
\label{Diff_trans} \end{equation}

where $z_{e} = 2l^{*}/3$ is the extrapolation length outside the
sample, at which the diffuse flux extrapolates to zero \cite{Morse and
Feshbach}. For $L < 3l^{*}$ the diffuse transmission probability
given by eq.(\ref{Diff_trans}), deviates significantly from experiment
\cite{Kaplan}. However, we choose to ignore this as we wish to obtain
estimates for imaging through thick scattering slabs where eq.(2) is
valid. Out of the fraction of the incident intensity that is
transmitted, a fraction ${\bf I_{diff}}$(L) is randomised and is
converted to a diffuse flux. The remaining fraction $1 - {\bf
I_{diff}}$(L) is the polarisation preserving component ${\bf
I_{pol}}$(L) responsible for the formation of the image. The quantity
${\bf \hat{I}_{diff}}$ is the diffuse flux that emerges normally from
the cuvette and is imaged onto the detector .The rest of the diffuse
flux which emerges at other angles is rejected by the aperture placed
in front of the CCD camera. Whether or not an image is detected
depends on two factors, the dynamic range of the detector, and the
relative intensities of ${\bf I_{pol}}$(L) and ${\bf \hat{I}_{diff}}$.

The image bearing intensity incident upon the CCD array is simply
${\bf I_{pol}}$(L)$T_{d}$ since this flux is assumed to emerge
normally from the sample. The diffuse flux ${\bf \hat{I}_{diff}}$
that emerges normal to the cuvette and is imaged onto the CCD is
controlled by the aperture {\bf PH}. It can be shown that for an
aperture of diameter $d_{a}$ placed in the focal plane of a lens of
focal length $f$, the angular acceptance of the lens is equal to the
angle subtended by the aperture as measured from the centre of the
lens \cite{Klein and Furtak}. In our experiment, an aperture with a
diameter of 1.3mm was placed at the focal plane of a lens with a focal
length of 90 mm. Thus, the acceptance angle $\delta$ of the limiting
aperture is $\sim$ 0.83 degrees.

To quantitatively determine the fractions of the various fluxes making
up the transmitted intensity, it only remains for us to estimate the
fraction of the diffuse intensity that is scattered out from the
cuvette within the acceptance angle $\delta$ of the lens aperture
system. We use a result due to Durian for the angular distribution of
diffusely transmitted light \cite{Durian3}. The expression obtained,
based on the diffusion approximation, for the probability for a photon
to be transmitted between the angles $\cos^{-1}(\mu)$ and
$\cos^{-1}(\mu + d\mu)$ from the exterior normal is given by

\begin{equation} P(\mu) = \frac{z_{e}\mu +
\mu^{2}}{\frac{z_{e}}{2}+\frac{1}{3}} \end{equation}

In deriving this expression, it has been assumed that there is no
refraction at the walls of the cuvette and that the boundary
reflectivity is independent of angle and polarisation. A more
detailed expression has also been derived \cite{Vera} which includes
these effects. However, since we are attempting only an
order-of-magnitude estimation, we omit these details. Integrating
$P(\mu)$ from $-\delta/2$ to $\delta/2$, we find the fraction of the
diffuse flux emerging within the acceptance angle of the lens aperture
system to be $\sim$ 0.007. However, since this randomly polarised
intensity has to pass through a polariser before reaching the CCD
array, only half the flux is transmitted resulting in ${\bf
\hat{I}_{diff}} \sim 3.5 \times 10^{-3} \cdot {\bf I_{diff}}$(L)$\cdot
T_{d}$. In summary, for a unit intensity incident on the scattering
medium, the total transmitted intensity $T_{d}$ may be partitioned as
follows :  (a) polarisation retaining fraction of the transmitted
intensity = ${\bf I_{pol}}$(L) $\cdot T_{d}$; (b) randomised fraction
of the transmitted intensity = ${\bf I_{diff}}$(L) $\cdot T_{d}$; and,
(c) fraction of the randomised transmitted intensity incident on the
detector = ${\bf \hat{I}_{diff}} \sim 3.5 \times 10^{-3} \cdot {\bf
I_{diff}}$(L)$\cdot T_{d}$. Figure 10 shows a plot of these fractions
for scatterers with $g$ = 0.45, the anisotropy of the particles used
in the experiment. Curve `a' is the diffuse transmitted fraction of
the total incident intensity $T_{d}$. Curves `b' and `c' are the
polarisation preserving intensity ${\bf I_{pol}}$(s), and the
randomised intensity selected by the aperture ${\bf\hat{I}_{diff}}$(s)
respectively, and $s$ is the distance travelled into the scattering
medium measured in units of the transport mean free path.

\begin{figure}[!htbp]
\centerline{\psfig{figure=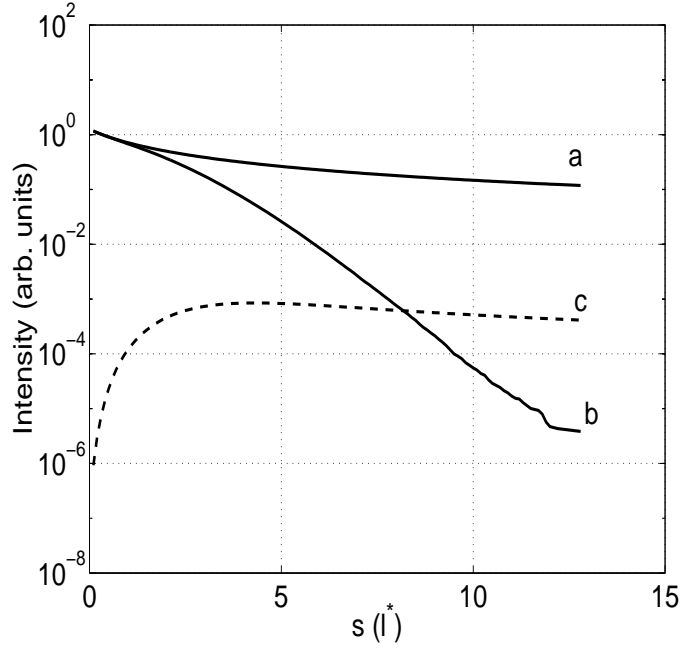,width=9cm,height=9cm}}
\caption{The total transmitted intensity $T_{d}$, the polarisation preserving
fraction ${\bf I_{pol}}$(s) $\cdot T_{d}$, and the normally emerging
diffuse component selected by the aperture ${\bf \hat{I}_{diff}}$, are
shown for varying slab thicknesses $s$. The curves are obtained for $g
= 0.45$, the anisotropy of the scatterers used in the experiment. The
fractions are calculated assuming that a beam of unit intensity is
incident on the medium. Curves `a', `b' and `c' represent $T_{d}$,
${\bf I_{pol}}$(s) $\cdot T_{d}$, and ${\bf \hat{I}_{diff}}$
respectively.}
\label{fluxes}
\end{figure}

Finally, we must obtain the criterion by which we can decide whether
an image may be extracted or not. When the polariser and analyser are
crossed, only the diffuse intensity is seen by the CCD and this is the
lowest intensity that will be observed. When they are parallel, the
CCD records the sum of the diffuse and polarisation preserving
intensities. Thus, the intensity time series for any pixel fluctuates
between these two values. However, with an 8 bit detection system
such as ours, the ripple in the intensity above the baseline due to
the rotating polariser, will not be detected when the amplitude of the
ripple is less than $\frac{1}{256}$ of the baseline intensity. In
practice though, for optically thick samples, the baseline is not
actually at zero when the polarisers are crossed because a large
diffuse flux is present and this further reduces the available dynamic
range of the detector. Therefore, the limit of detection in our
experiment is reached when ${\bf I_{pol}}$(s) $\le$
$\frac{1}{256}{\bf\hat{I}_{diff}}$(s), assuming that the entire
dynamic range is available at all $\tau$.

Figure 11 shows a semilogarithmic plot of the fluxes ${\bf
I_{pol}}$(s) and ${\bf\hat{I}_{diff}}$(s) for three values of $g$.
The curves marked `a', `b' and `c' refer to $g$ values of 0.1, 0.45
and 0.9 respectively. The dashed line represents the intensity level
below which the dynamic range of the detector is insufficient to
record an image. Curves `a' and `c' do not intersect with the dashed
line, while curve `b' comes very close to it. The reason is that the
largest simulation we can perform can propagate $10^{6}$ photons and
we have insufficient data at large depths to exactly determine the
point of intersection. However, the trends are clearly discerned.
Curve `b', which represents a $g$ value of 0.45, similar to our
experiment, is seen to intersect curve `d' at $\sim 13l^{*}$.
Experimentally, we find our limit of visibility to be at $\sim
10l^{*}$. Given that the dynamic range is greatly reduced at large
$\tau$ and that at high gain the detector is very sensitive to stray
light, this is in good agreement with our model.

We had previously commented, while discussing the results of our
simulations shown in Fig.9, on the rate at which the number of
polarisation preserving photons in the medium decreased with
increasing depth. It was seen surprisingly, that the input beam was
converted to a diffuse flux more rapidly with increasing $g$. We now
examine the consequences that this rate of decay has on the image
visibility assuming an imaging arrangement such as the one we have
used. We see interestingly that for $g = 0.9$, the curve steeply
drops towards the limit of visibility near $s = 7l^{*}$, while the
limit of visibility for $g = 0.45$ is approximately $13l^{*}$. For $g
= 0.1$, the curve flattens out and is nearly constant upto $s =
13l^{*}$. We have been unable to obtain data beyond this depth.
However, it indicates that both the depth upto which snake photon
imaging may be performed, and the quality of the image as determined
by the parameter {\bf R} are both significantly larger and better for
isotropic scatterers.

\newpage
\begin{figure}
\centerline{\psfig{figure=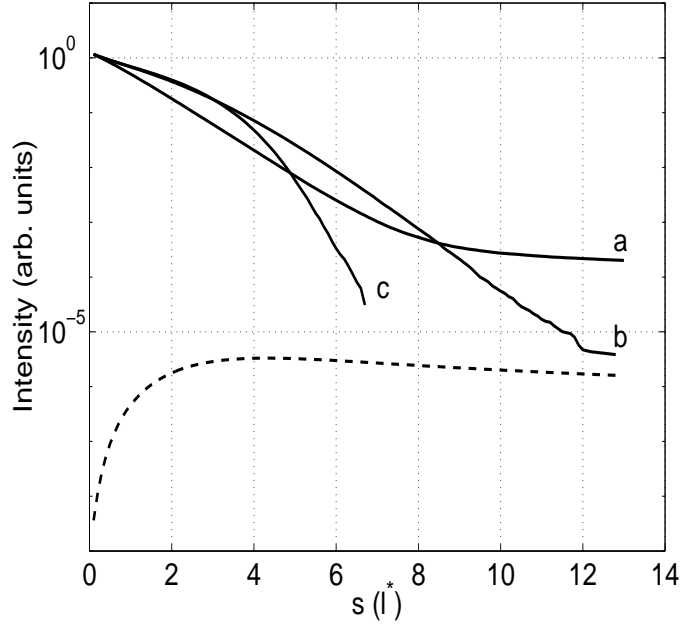,width=9cm,height=9cm}}
\caption{Semilogarithmic plot comparing the fraction of the transmitted intensit
y
${\bf I_{pol}}(s) \cdot T_{d}$, that still retains its original state of
polarisation after travelling a distance $s$ into the medium, for
different values of the scattering anisotropy $g$. The dashed line is
the is the limit of visibility set by the dynamic range of the detector
as $\frac{1}{256}{\bf\hat{I}_{diff}}$(s). The point of intersection of
the dashed line with any of the curves is the optical depth beyond which
snake photon imaging is not possible. The curves marked `a', `b' and
`c' represent scattering anisotropies of 0.1, 0.45 and 0.9 respectively.}
\label{visiblim}
\end{figure}
\subsection{Assumptions and approximations}

A crucial input to our estimation of the image visibility has been the
calculation of the fraction of the snake component that has been
randomised, and obtaining consequently, the fraction that is
polarisation preserving. In our simulation, for N photons that are
launched from the origin, a number $N_{diffuse}$ of those photons
undergo $N_{d}$ scattering events .The randomisation probability is
obtained by normalising the randomisation length histograms by
$N_{diffuse}$. However not all these photons that are launched {\it
and} undergo $N_{d}$ scattering events are transmitted. While the
simulation stops propagating these trajectories after $N_{d}$ events,
some of them are still backscattered. Therefore, our calculation of
the polarisation preserving transmitted intensity, ${\bf I_{pol}}$(L)
$\cdot T_{d}$, tacitly assumes that these photons are all transmitted,
an assumption that is not strictly valid. Yet, we find the agreement
with experiment to be good enough to permit us to make this
simplifying approximation. Further work is in progress to study in
greater detail the path length distribution of these randomised
photons to understand why this error is not significant.

Another bias that has been implicit in all our discussion so far has
been that we have not demanded accuracy of our results at low optical
densities. We feel that the analysis of the scattering of light in
optically thin slabs is simple since almost all the light is
transmitted virtually undeviated despite some scattering. Thus,
although the transmission coefficient $T_{d}$ is inaccurate at low
optical densities, and corrected versions are available for this
expression \cite{Durian2}, we have not used them, both to keep matters
simple as well as because our main focus is in the regime $L > 4l^{*}$
where the diffuse flux is large.

\section{Conclusions and Comments}

From experiments using particles with $g = 0.45$, we have found that
imaging with quasi-ballistic light is possible to depths greater than
that predicted by diffusion theory. To summarise our findings, we
begin with our results concerning the crossover from the
quasi-ballistic to the diffuse regime. We have classified as
`diffuse', a photon that has lost all memory of its initial direction
of propagation. This angular memory is quantified in terms of the
angular correlation function $C(j) = \langle {\bf \hat{n}}(0) \cdot
{\bf \hat{n}}(j) \rangle$. To our knowledge, such a definition of
what constitutes a diffuse photon has not been used before and we find
that it is highly instructive and offers greater insight into the way
photons are converted from quasi-ballistic to diffusive transport We
have calculated, as a function of the scattering anisotropy, the
number of scattering events $N_{d}$ for $C(j)$ to fall to zero. We
find that, contrary to the assumption implicit in diffusion theory,
that a photon undergoes approximately $n = l^{*}/l_{s}$ scattering
events before it is randomised, it actually takes many more scattering
events for the photon to be randomised, as can be seen from Table 1
and Fig. 8. However, calculating $N_{d}$ does not provide
information on the typical length scale on which angular correlation
is lost.

To understand the distribution of lengths over which photons are
directionally randomised, we have used random walk simulations and
obtained the path length distribution of the snake photons. We have
assumed, that until a photon has undergone $N_{d}$ scattering events,
it retains all memory of its initial polarisation. Drastic though
this may seem at first sight, it surprisingly explains our
experimental results very well. Further work is in progress to
calculate the pathlength distributions of these snake photons, with
polarisation taken into account, in order to understand this
approximation better.

We observe that it takes upto a thickness of about $6 - 8$ transport
mean free paths, depending on $g$, before practically the entire snake
component is randomised .Thus we are able to explain the observation
that diffusing wave spectroscopy, when used in the transmission
geometry, provides accurate results only when slab thicknesses exceed
this limit \cite{Pine, Durian1}. The distribution of the
randomisation lengths is the central result of our work. It shows
that image bearing photons are present in the medium to a depth much
greater than that predicted by diffusion theory where there is an
exponential conversion, on a length scale of the order of $l^{*}$, to
diffusive transport. Thus our simulations explain why the diffusion
approximation, which models the source of diffusing photons as a delta
function at a depth of $l^{*}$ inside the medium, is inaccurate for
slab thicknesses less than $6-8l^{*}$.

The abrupt reduction in ${\bf R}$ is now simple to explain and is
dependent on the dynamic range of the detector. In our experiment, we
use an 8-bit, 256 grayscale level CCD camera. So, as long as ${\bf
I_{pol}}(s)$ is greater than 256 times the diffuse intensity
${\bf\hat{I}_{diff}}$(s) incident on the detector, the ratio remains
at 0.5. As soon ${\bf I_{pol}}(s) \le {\bf\hat{I}_{diff}}$(s) an
exponential fall in visibility is observed. After the limit of
visibility is crossed, the detector has insufficient dynamic range to
distinguish the oscillating intensity due to the snake component from
the diffuse background. The detector sees a constant intensity and
$R$ is a constant once again.

Finally, we address our aim of placing limits on the depths to which
direct imaging into a turbid medium is possible. While we have
constantly used the word `imaging' so far, we are actually looking at
the transmission of photons through a random medium and the extent to
which they preserve their original directions of propagation. In, for
example, a medical diagnostic situation, the limits we have found
would strictly be valid only for making shadowgrams of inclusions
whose scattering characteristics are different from the bulk medium in
which they are embedded. However, multiple shadows used in
conjunction with standard inversion procedures such as the
backpropagation algorithm \cite{Kalpaxis} can be used to create three
dimensional `images' of inclusions in the medium. These too are not
strictly images, in that we cannot detect surface features on an
embedded object. We can only detect the outlines of the embedded
object. Our simulations show that for the limit of visibility,
assuming an experimental setup such as ours, maximum imaging depths
range from $\sim 7l^{*}$ for $g = 0.9$ to $\sim 13l^{*}$ for $g =
0.45$. For $g = 0.1$, our simulations are unable to determine the
limit of visibility. We expect that it lies significantly greater
than $13l^{*}$. In fact \cite{Ramachandran} suggests that it lies
beyond $30l^{*}$. We find this result both surprising and
counterintuitive that the maximum imaging depth decreases with
increasing scattering anisotropy. We do not yet understand the origin
of this effect. It should be noted that by reducing the size of the
aperture {\bf PH}, or choosing a detector with a greater dynamic
range, the limit of visibility may be extended. Before closing we
would like to comment that this last result is consistent with the
observations of Bizheva {\it et al.}  \cite{Bizheva} that were
published at the time that our manuscript was being prepared. Similar
to the work of Kaplan {\it et al.}\cite{Kaplan}, Bizheva {\it et al.}
also study the transition to DWS, but by means of an optical coherence
tomography technique. They also find that, "for smaller scattering
anisotropy diffusing light is detected after a greater number of
photon random walks in the sample", and "the transition to the light
diffusion regime occurs at shorter path lengths for media with higher
scattering anisotropy".

\begin{center}
{\Large Acknowledgements}
\end{center}
We thank the Supercomputer Education and Research Centre (SERC) at the
Indian Institute of Science for computational facilities. AKS thanks
the Raman Research Institute for a visiting professorship.
AKS and VG\footnote{Author for correspondence,
e-mail : vgopal@physics.iisc.ernet.in}
thank the Board of Research in Nuclear Sciences, India, for financial
assistance.

\end{document}